\documentstyle[12pt,aasms4]{article}  

\lefthead{Cavaliere, Menci \& Tozzi}
\righthead{The Luminosity Temperature Relation}

\begin{document}

\title{The Luminosity-Temperature Relation\\
    for Groups and Clusters of Galaxies}

\author{A. Cavaliere}
\affil{Astrofisica, Dip. Fisica, II Universit\`a di Roma,\\
via Ricerca Scientifica 1, 00133 Roma, Italy}

\author{N. Menci}
\affil{Osservatorio Astronomico di Roma
via Osservatorio, 00040 Monteporzio, Italy}

\and

\author{P. Tozzi}
\affil{ Astrofisica, Dip. Fisica, II Universit\`a di Roma,\\
via Ricerca Scientifica 1, 00133 Roma, Italy}

\begin{abstract}
We model the effects of shocks on the diffuse, X-ray emitting  
baryons in clusters of galaxies. Shocks separate 
 the infalling from the inner gas nearly at equilibrium, and 
dominate the compression and the density gradients of the latter in the 
dark-matter potential of the cluster. 
We find that, independently of the detailed shape of the potential, 
 the density gradient is steeper and the compression factor larger 
for the richer clusters. 
We show, considering the different merging histories, that in the 
hierarchical cosmogony the above effects lead, in X-rays, 
 to a luminosity-temperature relation $L\propto T^5$ at the scale of groups  
 which flattens down to $L\propto T^{3}$ for rich clusters 
 in accord with the observations, and then saturates toward
$L\propto T^2$ for higher 
 temperatures. From the merging histories we also compute statistical 
fluctuations of the $L-T$ correlation. 
\end{abstract}

\keywords{galaxies: clusters -- galaxies: intergalactic 
medium -- hydrodynamics}

\section{Introduction}

The X-ray emission from clusters of galaxies
 enables direct probing of 
gravitationally bound and virialized regions with virial radii $R_v$ 
of a few Mpcs, comprising 
total masses $M\sim 10^{15} M_{\odot}$ mostly in dark matter (DM hereafter).  

On the one hand, the X-ray temperature 
$T\propto GM/R_v$ measures the depth of the potential wells. 
On the other hand, the bolometric luminosity $L\propto 
 n^2\,R_X^3\,T^{1/2}$ emitted as thermal bremsstrahlung 
by the intracluster plasma measures the baryon number 
density $n$ within the volume $R_X^3$.  The $L-T$ relation constitutes a 
crucial {\it link} between the physics of the baryon component and the 
dynamical properties of the DM condensations. 

The simplest model describing the 
former holds $n$ to be proportional to the average DM 
density $\rho$, and $R_X$ to $R_v$, 
so that $ n \propto \rho\propto M/R_v^3$ obtains 
(self-similar model, hereafter SS, Kaiser \markcite{kais} 1986). 
If so, the luminosity 
would scale as $L\propto \rho^{1/2}\,T^2$, which is inconsistent with the 
observed correlation close to $L\propto T^3$ 
(Edge \& Stewart \markcite{edst} 1991; Mushotzky \markcite{mush} 1994; 
Tsuru et al. \markcite{tsur} 1996).  
Further steepening at the 
temperatures of galaxy groups is indicated for the emission not 
associated with  single galaxies (Ponman et al. \markcite{ponm} 1996). 
In addition, the SS model when combined with the standard 
hierarchical cosmogony (see Peebles \markcite{peeb} 1993) yields
for the clusters a local X-ray luminosity function too steep 
or too high compared with the data
(Evrard \& Henry \markcite{evhe} 1991; Oukbir, Bartlett \& Blanchard 
\markcite{oubl} 1996). 

So the indication is that the ratio $n/\rho$ 
is to depend on $M$ or $T$. 
It will be convenient to write the volume--averaged $n^2$ 
 in terms of two factors: the compression factor $g(T)\geq 1$, 
describing the gas overdensity relative to the outer value at the 
``boundary'', taken  here to be at $R_v$ as discussed in \S 3; 
and the inner shape factor $I(R_v,T)\equiv R_v^{-3}\int_0^{R_v}\,
d^3 {\bf{r}}\;  n^2 (r,T)/  n^2(R_v,T)$, with the 
main contribution coming from inside $R_X$.  
The result is:
\begin{equation}
L~ \propto ~g^2(T)\,I(R_v,T)\,R_v^3\,T^{1/2}\,\rho^{2}~\propto~
g^2(T)\;I(R_v,T)\;T^2\,\rho^{1/2}~, 
\end{equation}
where the last term follows from expressing $R_v$ from 
$T\propto M/R_v\propto \rho\,R_v^2$. 
 To obtain the {\it average} $L-T$ relation 
 to be compared with data, the 
 factor $g^2(T)$ in eq. (1) has to be 
 averaged over the cluster histories, as we carry out in \S  2.3. 

 So, the difference of eq. (1) from the SS model has been factored out into 
 the terms $g^2(T)$ and $I(R_v,T)$, which 
are determined by the hydro- and 
thermodynamics of the gas in the forming cluster wells. At $z\gtrsim 1$ 
the gas is expected to be {\it preheated} by 
feedback effects of star formation: 
injections of energy of stellar origin like Supernova winds 
eject the gas from the shallower potential wells, and 
 heat the residual and the 
ejected gas to temperatures $T_1\lesssim 10^7~K$ 
(Dekel \& Silk \markcite{desi} 1986; 
Kaiser \markcite{kais} 1991; Ciotti et al. \markcite{ciot} 1991; 
David et al. \markcite{dav1} 1993, 
\markcite{dav2} 1995; 
Cavaliere, Colafrancesco \& Menci \markcite{cacm} 1993). 
In addition, preheating is necessary 
to prevent too short cooling times (as pointed out 
 by Cole \markcite{cole} 1991, and addressed by 
 Blanchard, Valls Gabaud \& Mamon \markcite{bvgm} 1992; 
see also White \& Rees \markcite{whre} 1978) in early potential 
wells,  shallower on average but containing denser gas.  Subsequent 
evolution will lead to an increasing recovery of the universal baryonic 
fraction (White et al. 1993). 

Previous attempts (Kaiser \markcite{kais} 1991; Evrard \& Henry 
\markcite{evhe} 1991) 
to tackle the $L-T$ relation are based on the {\it extreme}
 assumption that
the gas inside the X-ray core $R_X$ 
is preheated but never subsequently 
shocked or mixed.
Here we discuss the {\it other} extreme, i.e., the 
effects of shocks and mixing on the 
gas density inside clusters.

\section{The Shock Model}

As in the collapses the gas velocity 
becomes supersonic, shock fronts form at about $R_v$, 
and separate the infalling from the inner gas already at virial 
temperatures.  In fact, numerical simulations (see Evrard 1990\markcite{evra}; 
Takizawa \& Mineshige \markcite{tami} 1997) of isotropic collapses 
show that, when the outer gas temperature is appreciably 
lower than the virial value $T$, a spherical 
shock front forms and, in the vicinity of $R_v$, slowly expands outwards 
leaving the gas nearly at rest and with a nearly flat 
 temperature profile.  
When realistic, {\it anisotropic} collapses are considered 
(Navarro et al. \markcite{nava} 1996, Tormen \markcite{torm} 1996) 
shock fronts still form and
convert into heat most of the hydrodynamical energy 
(Schindler \& M\"uller \markcite{shmu} 1993; Schindler \& B\"ohringer 
\markcite{shbo} 1993; R\"ottiger, Burnes,  \& Loken \markcite{rott} 1993).

Across the shock the gas entropy rises, and correspondingly 
a jump in the gas density and in the temperature 
is established 
 from the exterior values $ n_1$, $T_1$ to the interior 
 ones $ n_2$, $T_2$.  
The density jump provides a {\it boundary}  
condition for the inner gas distribution in the form of the 
compression factor $g(T_2/T_1)\equiv  n_2/  n_1$. In addition, the 
interior  temperature $T_2$ governs, at equilibrium 
in a given gravitational potential, the inner density {\it profile}  
 and hence the shape factor $I(R_v,T)$. 
These two effects enter eq. (1) for $L$, 
and  will be discussed in turn. 

\subsection{ The Compression Factor}

The values of $ n_2$, $T_2$ and of the interior gas velocity 
$v_2$ are related to their outer counterparts 
$ n_1$, $T_1$ and $v_1$ by the requirements of 
mass, momentum and energy conservation across the shock. 
The plasma behaves as a perfect gas with three degrees of freedom, 
and the Hugoniot adiabat (see Landau \& Lifshitz \markcite{lali}1959) 
yields the compression factor 
\begin{equation}
g\Big({T_2\over T_1}\Big) = 
2\,\Big(1-{T_1\over T_2}\Big)+\Big[4\, 
\Big(1-{T_1\over T_2}\Big)^2 + {T_1\over T_2}\Big]^{1/2}~.
\end{equation}
For strong shocks with $T_2\gg T_1$, this 
saturates to the value $g =4$, while for $T_2\rightarrow 
T_1$ it attains its lowest value $g=1$.  

The pre-shock temperature $T_1$ is provided by 
the stellar (thermonuclear) energy feedbacks 
recalled in \S 1, or by the virial (gravitational) 
temperature inside the clumps which are to 
merge with the cluster during its merging history considered below. 
In fact, the stellar feedbacks set for $T_1$ the lower bound 
$T_{1*}$ which 
we identify with the lowest temperatures 
(around $0.5$ keV, Ponman et al. 1996) measured in groups. 

Pre-shock temperatures of this order do not 
affect the rich clusters; instead, they affect the compression factor 
in the shallower potential wells with $T\sim 1$ keV, 
 as prevail at redshifts $z\gtrsim 1$ but are 
also present at $z\simeq 0$.  The full behavior 
 of $g(T_2/T_1)$ in shown by the dashed line in fig. 1 when 
 $T_1=T_{1*}$ and the latter is 
in the range $0.5-0.8$ keV. The actual values of $T_1$ 
will be discussed in  \S 2.3, taking into account the merging histories. 


\subsection{ The Gas Disposition}

The post-shock temperature $T_2$ can be calculated from the pre-shock velocity 
 $v_1$. The latter is driven  
by the gravitational potential $V(r)$, and reads 
$v_1 = [-\alpha\,V(R_v)/m_p]^{1/2}$ 
with $\alpha = 2[1-V(R_{m})/V(R_v)]$.  Here  
 $R_{m}$ is the radius where infall becomes nearly free;
its upper bound is obtained by equating the Hubble flow to the 
free-fall velocity, which yields $\alpha\approx 1.4$. 
The post-shock condition is closely hydrostatic, i.e., 
 $v_2 \ll v_1$, as shown by the simulations. 
Then, using the equations in Landau \& Lifshitz \markcite{lali} (1959)
with the above value of $v_1$, we find 
\begin{equation}
 kT_2\simeq -{\alpha \over 3}\,V(R_v)+ {7\over 8}\,kT_1~.  
\end{equation}

The inner gas profile relative to that of DM, 
$\rho (r)$ say, is governed at equilibrium by the 
scale-height ratio $\beta\equiv \mu\,m_p\sigma_r^2/kT_2$, 
where $\mu\approx 0.6$ is the gas mean molecular weight, $m_p$ 
the proton mass, and 
$\sigma_r$ the one--dimensional velocity dispersion of the DM. The profile 
\begin{equation}
n(r) \propto [\rho(r)]^{\beta(T)} 
\end{equation}
(Cavaliere \& Fusco-Femiano 1976), applies 
for a nearly flat $T(r)\approx $ const $\approx T_2$. 

The function $\beta (T)$ entering the profile (4) 
is easily computed from eq. (3), for 
a given DM potential V(r) corresponding to $\rho(r)$.  
For the King potential (see Sarazin \markcite{sara} 1988)  
 $V(r)=-9\, \mu m_p\, \sigma_r^2r_c\,ln[(r/r_c)+(1+r^2/r_c^2)^{1/2}]/r$ 
with the core radius $r_c = R_v/12$, we obtain 
  $\beta(T)$ increasing somewhat from the value 
$\beta\approx 0.5$ for $T\approx T_1$ to 
$\beta \approx 0.9$ for $T\gg T_1$.  
A similar result obtains using the potential 
proposed by Navarro et al. \markcite{nava} (1996).  
These two instances are illustrated in fig. 2, and demonstrate 
that in all cases the static gas density profile is {\it shallower} for 
lower $T$ clusters. 


\subsection{$L-T$ from Merging Histories}

For a given $T_1$ the compression factor $g$ is computed after eq. (2), and 
 the shape factor $I(R,T)$ is obtained by integration of $ n^2(r)$ 
 computed after eq. (4); then the $L-T$ relation may be obtained from eq. 
(1). But $T_1$ depends on the thermal conditions of the infalling gas, which 
is preheated by Supernovae and {\it further} heated 
 through virialization inside merging clumps. 

In the former case, we take $T_1=T_{1*}=0.5 - 0.8$ keV with a flat 
distribution, and obtain for  $g^2$ the 
dashed line in fig. 1.
In the latter case, repeated merging events introduce fluctuations of 
  $T_1$ above $T_{1*}$, 
and hence of the interior density $n_2$, which modify $g^2$. 
The average effect 
 is shown by the solid line in fig. 1, and the corresponding variance is 
illustrated by the shaded area. 

To include both conditions, we take 
 $T_1$ to be the {\it higher} between the preheating value $T_{1*}$ and 
the virial temperatures  prevailing 
in the clumps accreted by the cluster. We perform 
a statistical convolution 
of the $L-T$ relation over such merging histories, using Monte Carlo 
realizations of hierarchical merging trees of 
dark halos as introduced by Cole \markcite{cole} (1991). 
The code, written by one of us (P.T.), is based 
directly on the excursion set approach of Bond et al. \markcite{bond} (1991) 
to the mass distribution.  

For each merging event concerning a cluster with a current virial  
temperature $T'$, we compute $g(T'/T_1)$ for the clumps being accreted,  
weighted with the associated mass fraction. We show in fig. 1 the quantity 
$\langle g^2\rangle$ averaged over the histories ending into a 
cluster of temperature $T$, along with its dispersion. These two quantities  
 are used in eq. (1) to predict the average $L-T$ correlation and its 
scatter. 
The results are shown in fig. 3, for the 
 DM potential of Navarro et al. \markcite{nava} 1996; a different $V(r)$, like
King's, only steepens somewhat the low-$T$ behavior. 


\section{Results and Discussion}

Here we have proposed a model for the intracluster gas, to capture in a simple
way one essential component of the complex gravitational systems   
constituted by groups and clusters. 
The model is focused on the formation of {\it shocks} 
between the the gas nearly in equilibrium with the cluster potential, 
and the 
infalling one. The latter is preheated by the release of 
thermonuclear energy, or by the gravitational energy in subclusters. 
Shock heating and compression determine the $L-T$ relation. 

The latter comprises both clusters and 
small groups in a single dependence, which smoothly {\it flattens} 
 from $L\propto T^5$ (for groups with $T\lesssim 2$ keV), to 
$L\propto T^{3.5}$ (for clusters with 2 keV$\lesssim T\lesssim$7 keV), toward
$L\propto T^2$ (for higher $T$). Such behavior fits 
both the cluster data (Edge \& Stewart \markcite{edst} 1991) and those 
for groups (Ponman et al. \markcite{ponm} 1996) 
which --  if considered separately -- 
 would require a much steeper $L-T$ relation. 
Correspondingly, the volume-averaged baryonic fraction grows 
by a factor around 3 from small groups to rich clusters, but remains 
 within 1.3 times the universal value. 

In addition, we predict the gas
density profiles to have a flat central region (the gas ``core'', see fig. 2), 
independently of the detailed shape of the 
DM potential; the profiles are actually 
{\it steeper} for larger virial 
temperatures. 
Correspondingly, the size $R_X$ of the X-ray core (defined, e.g., 
at one half the integrated emission)   
grows with mass slower than $M^{1/3}$. The average cooling time within $R_X$ 
exceedes the Hubble time out to $z\approx 2$, differently from the SS model.  

We have checked that these results {\it persist} when we relax the 
approximation 
of a flat temperature profile in the cluster, and adopt instead a polytropic
distribution with index $\gamma$ ranging from $1$ to $5/3$; for 
 For $\gamma >1$ the temperature declines from the center toward the shock 
position. 

In time, the shocked region expands and outgrows $R_v$, 
 the infall velocity decreases, and 
the shock weakens with $T_2$ approaching $T_1$. 
 However, this occurs only over several dynamical times  as  shown by the 
 N-body simulations (see Takizawa \& Mineshige 1997\markcite{tami}); 
 meanwhile, the shock positions remain close to $R_v$, 
as taken here. 

We do not stress, instead, the $z$-dependence $(1+z)^{1.5}$  
of the normalization provided by the factor $\rho ^{1/2}(z)$ 
appearing in eq. (1). 
In fact, such dependence is easily swamped by the place-to-place 
variations of $n_1$  
and by the systematic increase in contrast of the large scale 
structures hosting groups and clusters (see Ramella, Geller, \& Huchra 
1992\markcite{rame}). 

The robust predictions of the shock model do not require spherical 
symmetry, but only a small bulk velocity of the inner  gas 
compared to $(GM/m_p\, R_v)^{1/2}$. 
Thus the model includes {\it anisotropic}, 
recurrent merging with other clumps of dark and baryonic matter. 

The effects of extreme merging events are as follows. 
The few events involving {\it comparable} subclusters reshuffle 
the baryonic content and mix its entropy, 
but only moderatly affect temperature and density.  
At the other extreme, the more isotropic accretion to a cluster of many 
{\it small} condensations {\it gravitationally} heated at 
 temperatures $T_1\approx 1$ keV 
yields the highest compression and the largest contribution to the X-ray 
luminosity. 
Our {\it Monte Carlo} simulations span the range between these extremes. 

The stellar preheating at $T_{1*}$ of the external gas is essential  
 to provide  a lower limit for $T_1$. This ensures 
that the accreting gas, even when in a very shallow well or in a diffuse 
 state, starts on a relatively high adiabat, corresponding to 
$T_{1*}=0.5 - 0.8$ keV. 
Such a heating of {\it thermonuclear} origin breaks down the otherwise 
self-similar form $n/\rho=$cost of the ratio of gas to 
DM density to yield, at the boundary, 
the form $n/\rho\propto g(T/T_1)$ shown in fig. 1. 
 Values of $T_{1*}$ smaller than 0.5 keV would lead, in a strictly self-similar 
evolution of DM halos, to 
$L\propto T^2$ at variance with the data; 
larger values to a severe depletion of the gas and of the luminosities 
in groups and clusters. 

The merging histories also produce the considerable scatter in the $L-T$ 
relation shown in fig.3, since the different virial temperatures of 
the stochastically merging clumps induce intrinsic variance in the internal 
 density $n_2$.  
Further scatter may be contributed by the vagaries in the 
 ambient density $n_1$, and by the possible lack of 
 dynamical equilibrium in some groups, 
see discussion by Governato, Tozzi \& Cavaliere \markcite{cagt} 1996.

The shock model in the simple form presented here applies to
the gas settled to equilibrium after 
each dynamical perturbation. This takes sound propagation times, somewhat 
shorter than the dynamical timescale taken anyway by the DM to adjust 
to equilibrium (Tormen \markcite{torm} 1996). The residual converging motions, 
even in spherical N-body simulations (Takizawa \& Mineshige \markcite{tami} 
1997), tend to balance  the expansion of the 
shocked region to yield only small net velocities  $v_2\approx 100$ km/s. 
These may be associated with some adiabatic compression of the central regions, 
but the resulting heating is only mild, as long as shocks form 
at radii of order $R_v$. 

The other extreme is 
tackled by the model proposed by Kaiser \markcite{kais} (1991) 
and refined by Evrard \& Henry \markcite{evhe} (1991). 
This assumes that, after preheating, the central cluster 
region visible in X-rays contains 
the same gas (about 10 \% of the present total) engaged
in a smooth adiabatic compression. 
However, we find only $5\%$ of the largest single progenitors to have 
masses (DM and hence gas) exceeding $10\%$
of the present values at $z\geq 1.5$, when most stellar preheating
takes place.  On the other hand, N-body simulations (see refs. in \S 2) 
and observations  (see Zabludoff \& Zaritsky \markcite{zaza} 1995) show that 
each cluster history includes a few 
merging events between comparable structures, which will reset 
the core gas to a higher adiabat. 

The adiabatic model (with preheating) 
predicts a single, scale-free relation 
$L\propto T^{3.5}$, or $L\propto T^3$ if the gas equilibrium holds out to $R_v$ 
as in Evrard \& Henry (1991).  However, on the largest scales 
that ought to sample fairly the universal baryonic fraction 
(White et al. \markcite{wh93} 1993) one 
expects saturation toward  the scaling $L\propto T^2$ of the 
SS model; in addition , at the group scales a much 
steeper dependence is indicated by 
observations. Such opposite departures of
the $L-T$ correlation from a single power-law are beyond the reach 
of the adiabatic model, but within the predictions of 
 the shock model. 

The model we propose leads (Cavaliere, Menci, Tozzi \markcite{camt} 1997)  
to a local luminosity
function $N(L, z=0)$ in agreement with the data, 
and to $N(L,z)$  with the mild or no evolution
shown by recent data, and confirmed by the deep X-ray counts.  

\acknowledgments
We acknowledge informative discussions with M. Ramella, the helpful comments 
of the referee, and  grants from MURST and ASI. 

\clearpage

%
%

\clearpage

\figcaption[fig1.eps]{
The square compression factor $g^2(T/T_1)$ is shown by the dashed 
line when $T_1$ takes on its lowest value $T_{1*}$, with the latter 
uniformly distributed in the range $0.5-0.8$ keV. 
The different values of $T_1$ due to the 
merging histories (discussed in \S 2.3) affect the average dependence as
 shown by the solid line, and provide the 2-$\sigma$ 
dispersion shown by the shaded region. A tilted CDM power spectrum in a
critical cosmology (see White et al. 1996) has been used. 
\label{fig1}}

\figcaption[fig2.eps]{
The gas density profile (for a uniform $T$) 
derived from eqs. (3) and (4) 
using the King DM potential (upper panel), or (lower panel) that given by 
Navarro et al. (1996). 
The dotted lines refer to a group with 
$T=T_1=0.8$ keV, and the solid lines to a rich cluster with $T_1\ll T=8$ 
keV.  \label{fig2}}

\figcaption[fig3.eps]{
The L-T relation from the shock model, convolved with the merging histories 
of DM halos, is compared with data for clusters of galaxies 
(filled squares, from Edge \& Stewart 1991), and for groups 
(open squares, from Ponman et al. 1996). The DM potential of 
Navarro et al. (1996) is used. 
The shaded region outlines the $2-\sigma$ scatter expected from the merging 
histories. 
\label{fig3}}

\begin{references}

\reference{bvgm} Blanchard, A., Valls-Gabaud, D., \& Mamon, G. 1992, 
\aap, 264, 365
\reference{bcek} Bond, J.R., Cole, S., Efstathiou, G., 
\& Kaiser, N. 1991, \apj,  379, 440
\reference{cacm} Cavaliere, A., Colafrancesco, S., 
\& Menci, N. 1993, \apj, 415, 50
\reference{cafu} Cavaliere, A, \& Fusco Femiano, R. 1976, \aap, 49,137
\reference{camt} Cavaliere, A., Menci, N., \& Tozzi, P. 1997,  in preparation
\reference{ciot} Ciotti, L., D'Ercole, A., Pellegrini, S., \& Renzini, A. 
1991, \apj, 376, 380
\reference{cole} Cole, S. 1991, \apj,  367, 45
\reference{dav1} David, L.P., Slyz, A., Jones, C., Forman, W., Vrtilek, 
 S.D., \& Arnaud, K.A. 1993, \apj, 412, 479
\reference{dav2} David, L.P., Jones, C., \& Forman, W. 1995, \apj, 445, 578
\reference{desi} Dekel, A., \& Silk, J. 1986, \apj, 303, 39
\reference{edst} Edge, A.C., \& Stewart, G.C. 1991, \mnras, 252, 414
\reference{evra} Evrard, A.E. 1990, \apj, 363, 349
\reference{evhe} Evrard, A.E., Henry, J.P. 1991, \apj, 383, 95
\reference{cagt} Governato, F., Tozzi, P., \& Cavaliere, A. 1996, \apj, 458, 18
\reference{kais} Kaiser, N. 1991, \apj, 383, 104
\reference{lali} Landau, L.D., Lifshitz, E.M. 1959, {\it Fluid Mechanics} 
(London: Pergamon Press), p. 329-331
\reference{mush} Mushotzky, R.F. 1994, in {\it Clusters of Galaxies}, 
eds. F. Durret, A. Mazure, Tran Tahn Van J. (Gif-sur-Yvette: 
Ed. Fronti\`eres), 177
\reference{nava} Navarro, J.F., Frenk, C.S., \& White, S.D.M. 1996, preprint 
[astro-ph/9611107] 
\reference{oubl} Oukbir, J., Bartlett, J.G., Blanchard, A
. 1996, preprint [astro-ph/9611089]
\reference{peeb} Peebles, P.J.E. 1993, 
{\it Principles of Physical Cosmology} (Princeton: Princeton Univ. Press)  
\reference{ponm} Ponman, T.J., Bourner, P.D.J., Ebeling, H., B\"ohringer, 
H. 1996, preprint
\reference{rame} Ramella, M., Geller, M.J., \& Huchra, J.P. 1992, \apj, 384, 
396
\reference{rott} R\"ottiger, K., Burnes, J., Loken, C. 1993, \apj, 457, L53
\reference{sara} Sarazin, C.L. 1988, {\it X-ray Emission from Clusters of 
Galaxies} (Cambridge: Cambridge University Press)
\reference{shbo} Schindler, S., B\"ohringer, H. 1993, \aap, 269, 83
\reference{shmu} Schindler, S., \& M\"uller, E. 1993, \aap, 272 137
\reference{tami} Takizawa, M., \& Mineshige, S. 
1997, preprint [astro/ph 9702047]
\reference{torm} Tormen, G. 1996, \mnras, in press [astro-ph/9611078]
\reference{tsur} Tsuru, T. et al.  
1996, in {\it The 11th Int. Coll. on UV and X-ray 
Spectroscopy of Astrophysical and Laboratory Plasmas}, eds. Watanabe, T., 
Yamashita, K., in press
\reference{wh96} White, M., Viana, T.P., Liddle, A.R., 
\& Scott, D. 1996, \mnras, 283, 107
\reference{whre} White, S.D.M., \& Rees, M. 1978, \mnras, 183, 341
\reference{wh93} White, S.D.M., Navarro, J.F., Evrard, A.E., \& Frenk, 
C.S. 1993, \nat, 366, 429 
\reference{zaza} Zabludoff, H.I. \& Zaritsky, D. 1995, \apj, 447, L21

\end{references}
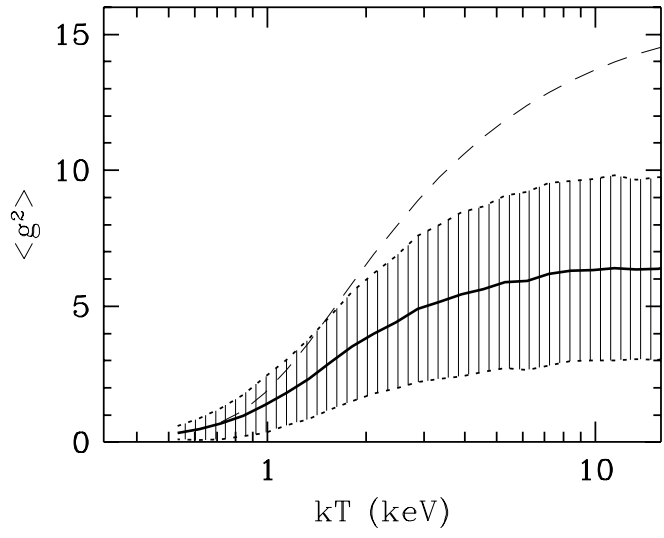
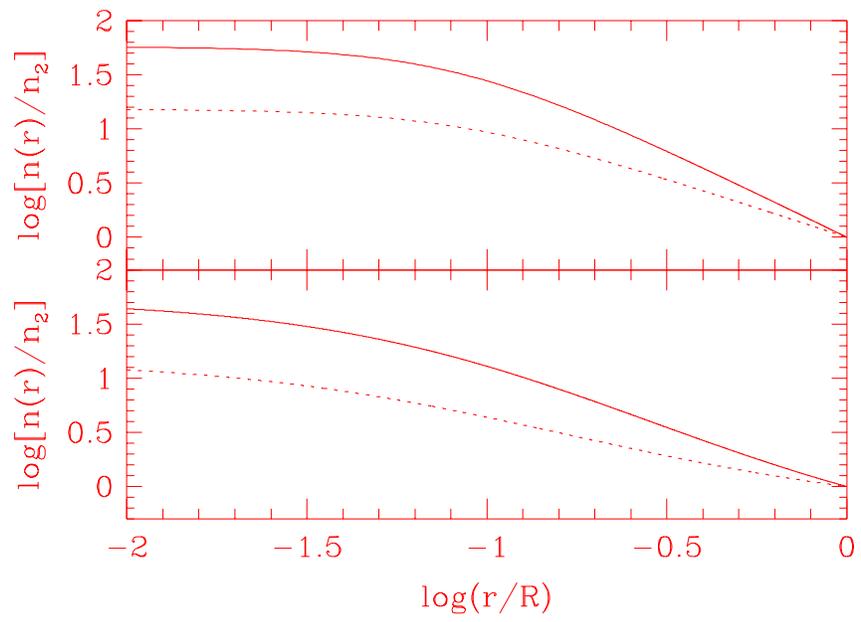
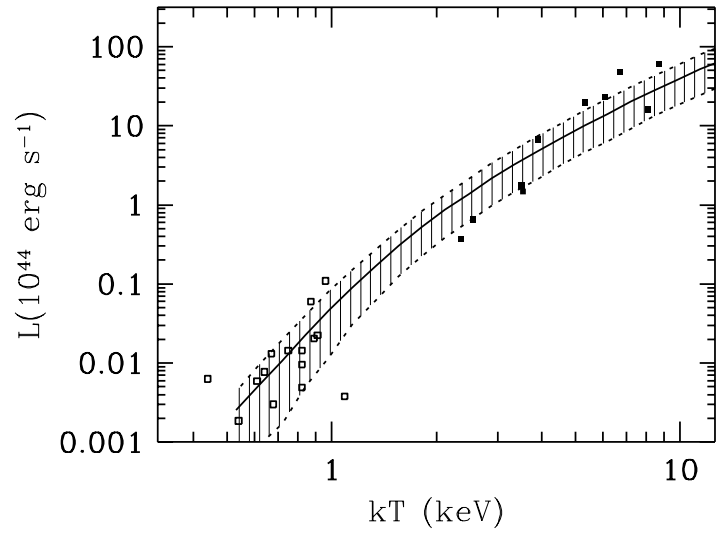
\end{document}